\documentclass[prl,twocolumn,showpacs,amsmath,amssymb]{revtex4-1}

\usepackage{calc}
\usepackage{float}
\usepackage{graphicx}% Include figure files
\usepackage{bm}% bold math
\usepackage[percent]{overpic} %for inset figures NK
\usepackage[colorlinks=true]{hyperref} %for hyperlinking NK
\usepackage{amsmath}
\usepackage{amssymb}
\usepackage{latexsym}
\usepackage{multirow}

\newcommand{\ket}[1]{\left| #1 \right>} % for Dirac bras
\newcommand{\bra}[1]{\left< #1 \right|} % for Dirac kets
\newcommand{\matrixel}[3]{\left< #1 \vphantom{#2#3} \right|
	#2 \left| #3 \vphantom{#1#2} \right>} % for Dirac matrix elements
\newcommand{\abs}[1]{\left|\,#1\,\right|} % for absolute value
\newcommand{\avg}[2]{\left<\, #1 \,\right>_{#2}} % for average
\begin{document}
\title{Universality at Breakdown of Quantum Transport on Complex Networks}
\author{Nikolaj Kulvelis}
\author{Maxim Dolgushev}
\author{Oliver M\"ulken}
\email{muelken@physik.uni-freiburg.de}
\affiliation{Physikalisches Institut, Universit\"at Freiburg, Hermann-Herder-Str.\ 3, D-79104 Freiburg, Germany}

\date{\today} 
\begin{abstract} 
We consider single-particle quantum transport on parametrized complex networks. Based on general arguments regarding the spectrum of the corresponding Hamiltonian, we derive bounds for a measure of the global transport efficiency defined by the time-averaged return probability. For tree-like networks, we show analytically that a transition from efficient to inefficient transport occurs depending on the (average) functionality of the nodes of the network. In the infinite system size limit, this transition can be characterized by an exponent which is universal for all tree-like networks. Our findings are corroborated by analytic results for specific deterministic networks, dendrimers and Viscek fractals, and by Monte Carlo simulations of iteratively built scale-free trees.
\end{abstract}

\maketitle
\date{\currenttime}

%\section
{\sl Introduction} -- 
Complex networks are intensively used as models for a panoply of physical, chemical, biological, or sociological systems, see \cite{barabasi2009scale,newman2010networks, barthelemy2011spatial} and references therein. They have been proven to be extremely useful in understanding the statistical as well as the dynamcial features of these systems, e.g., the complexity of the structure of the internet is captured by scale-free networks, where the probability distribution of the number of connections of a given node (hub) follows a power law \cite{albert1999internet}. In contrast to stochastic networks, also deterministic networks - such as regular, hyper-branched, or fractal types - have been applied to study the properties of, say, macromolecular compounds and polymers \cite{Gurtovenko2005,Reuveni2010}. 

Interestingly, many topologically different networks show similar statistical (equilibrium) properties, i.e., they fall into the same universality class \cite{barthelemy2011spatial}. Recently, it has been shown that, in addition to these classes, there are also dynamic universality classes, referring to similar classical diffusive dynamics' of different type and on different networks \cite{barzel2013universality}. 
However, it is by no means clear that a corresponding quantum dynamics on such networks allows for a classification based on a quantum analog of dynamic universality. As will be presented below, one can define such a quantum analog also for (single-particle) quantum transport processes on complex networks. In particular, in this paper we will be concerned with tree-like networks, i.e., with networks without loops. For these we find that one can indeed define classes which solely depend on the spectrum of the corresponding Hamiltonian. As we show, classes of parametrized Hamiltonians lead to a power-law dependence of a (time-averaged) characteristic quantity describing the transport efficiency \cite{mulken2011continuous}.

%\section
{\sl Quantum Transport Efficiency} -- 
We consider 
closed, microscopic physical system modelled by undirected graphs/networks $ \mathcal{G}(\mathcal{V},\mathcal{E})$ having a set $\mathcal{V}$ of vertices/nodes and a set $\mathcal{E}$ of edges/bonds. To each node $j$, out of the total number of $N$, we associate a (basis) state $\ket{j}$, such that all these states form a basis set of the Hilbert space. The direct couplings between two nodes $j$ and $k$ are mediated by a bond; the number of direct couplings of node $j$ is called functionality (or degree) $f_j$. The dynamics starting from a localized state $\ket{j}$ is then modelled by Schr\"odinger's equation ($\hbar\equiv 1$) for the transiton amplitudes $\alpha_{k,j}(t) \equiv \bra{k} \exp(-i \bm H t) \ket{j}$, which is properly solved by diagonalizing $\bm H$ \cite{farhi1998quantum,mulken2011continuous}.
We will consider only such Hamiltonian with indentical coupling strengths $H_{k,j} \equiv \matrixel{k}{\bm H}{j} = 1$ between any pair of nodes connected by a single bond and with \emph{on-site} potentials $H_{j,j} = H(f_j)$, i.e., nodes of the same functionality have the same potential. 
Matrices typically associated with graphs $\mathcal{G} $, like its connectivity matrix ($H(f_j)=f_j$) and its adjacency matrix ($H(f_j)=0$ for all $j$), are contained in this class.

Based on the transition amplitudes $\alpha_{k,j}(t)$ and the corresponding transition probabilities $\pi_{k,j}(t)=|\alpha_{k,j}(t)|^2$, we will characterize a network's transport efficiency by the average return values
$\bar{\alpha}(t)=\tfrac{1}{N}\sum_{j=1}^{N}\alpha_{jj}(t)$ and  $\bar{\pi}(t)=\tfrac{1}{N}\sum_{j=1}^{N}\pi_{jj}(t)$
 \cite{mulken2011continuous}.
Generally, values of $\bar{\pi}(t)=\mathcal{O}(1)$ for almost all $t$ imply - on average - a high probability for an excitation to remain at the initial node, thus indicate inefficient transport, values of $\bar{\pi}(t)\ll 1$ for almost all $t$ suggesting the opposite. Clearly, one  needs the entire knowledge of $\bm H$'s eigenspace. However, for $\bar{\pi}(t)$ this can be circumvented by using the \textsc{Cauchy-Schwarz} inequality to obtain a lower bound \cite{mulken2011continuous}
\begin{equation}
	\label{eq:alpha_pi_ineq_spec}
	\abs{\bar{\alpha}(t)}^2 = \sum_{E,E'}\varrho(E)\varrho(E')e^{-i(E-E')t} \leq \bar{\pi}(t)\;,
\end{equation}
where $\abs{\bar{\alpha}(t)}^2$ solely depends on the (discrete) spectral density $\varrho(E)$ of $\bm H$.
An asymptotic time-independent measure for the global transport efficiency can be introduced by the return-quantities's infinite time limit
\begin{equation}
	\label{eq:chi_def}
	\chi\equiv\lim_{t\to\infty}\frac{1}{t}\int_{0}^{t}\text{d}t'\abs{\bar{\alpha}(t')}^2\leq\lim_{t\to\infty}\frac{1}{t}\int_{0}^{t}\text{d}t'\bar{\pi}(t')\;.
\end{equation}
As a lower bound, $\chi$ is most instructive if $\chi=\mathcal{O}(1)$, thus the RHS of Eq.~(\ref{eq:chi_def}) is also of order $\mathcal{O}(1)$. Then we regard the global transport as being inefficient, whereas for $\chi\ll\mathcal{O}(1)$ there is no \emph{strict} implication on the exact value of the infinite time limit of $\bar\pi(t)$. However, previous results suggest that in particular the maxima of $\bar\pi(t)$ are well reproduced by the lower bound $|\bar\alpha(t)|^2$, therefore, also indicating that the values of the RHS of Eq.~(\ref{eq:chi_def}) lie close to the values of $\chi$ \cite{mulken2011continuous}.

Combining Eqs.~\eqref{eq:alpha_pi_ineq_spec} and \eqref{eq:chi_def} allows to estimate $\chi$ from below, knowing only the spectral density $\varrho(E_{*}) $ of one arbitrary eigenvalue $E_{*}$, 
\begin{equation}
	\label{eq:chi_spec_dens_ineq}
	\chi=\sum_{E}\varrho^2(E)\geq \varrho^2(E_{*})+\frac{1}{N}\big[1-\varrho(E_{*})\big]=\underline{\chi}\;.
\end{equation}
Here, we assume a completely flat density $ \varrho(E) $ on its support aside from $E_{*}$. $\underline\chi$ allows for a rather accurate extimation of $\chi$ if $E_*$ is a single highly degenerate eigenvalue compared to all other eigenvalues, i.e. if $\varrho(E_{*})\gg\varrho(E\neq E_{*})$, or if all eigenvalues are non-degenerate, i.e., $\rho(E)=1/N$ for all $E$. These two limits correspond to vastly different networks. For instance, chain-like networks with $\bm H$ belonging to the above mentioned class have eigenvalues with (mostly) the same degeneracy, i.e., $\rho(E) = \text{const.}$ for all $E$, which yields $\underline\chi=\chi=1/N$ \cite{mulken2011continuous}, suggesting very efficient transport. In contrast to this, stars of the same size typically have a single highly degenerate eigenvalue yielding $\underline{\chi}=\chi=1-(4N-6)/N^{2}$ \cite{mulken2011continuous}, thus, rendering transport inefficient. Obviously, in the infinite system size limit one has 
\begin{equation}
\underline\chi_\infty \equiv \lim_{N\to\infty} \underline\chi =
\begin{cases}
0 \qquad \text{for chains}\\
1 \qquad \text{for stars}
\end{cases}\;.
\end{equation}

%\section
{\sl Breakdown of Quantum Transport} -- 
For networks whose spectra dependent on a tunable paramter $\sigma\in\mathbb{R}$ such that for large values of $\sigma$ the spectral density of the  selected eigenvalue $E_*$ is of order $\mathcal{O}(1/N)$ and for small values of $\sigma$ of order $\mathcal{O}(1)$, one might observe a transition from efficient to inefficient transport. Let this transition occur at a given parameter value $\sigma_c$. 
At this value the transport breaks down which is reflected by a change of $\underline\chi(\sigma)$ from values of $\mathcal{O}(1/N)$ for $\sigma>\sigma_c$ to values of $\mathcal{O}(1)$ for $\sigma<\sigma_c$. This is reminiscent of a \emph{phase transition} where, in our case, the quantity $1-\underline\chi_\infty(\sigma)$ represents the \emph{order parameter}.
(We use the usual terminology of phase transitions in order to stress the similarities and to avoid to overload the paper with new terminology.)
Consequently, we associate with this transition a \emph{critical exponent} defined by 
\begin{equation}
\underline\kappa\equiv\lim_{\sigma\to\sigma_{c}} \frac{\log |1-\underline\chi_\infty(\sigma)|}{\log \abs{\sigma - \sigma_{c}}}.
\end{equation}
Since $\underline\chi_\infty(\sigma)\leq\chi_\infty(\sigma)$, one has $\underline{\kappa}\geq\kappa $, where $\kappa$ is the exponent associated with the transition for $\chi_\infty(\sigma)$. As we will show below, tree-like networks which have a parametrized transition from chain-like topologies to star-like topologies yield the same exponent $\underline\kappa$. This allows us to group networks, depending to their asymptotic (global) quantum transport efficiency, into \emph{universal classes} defined by $\underline\kappa$.

\begin{figure}[H]
\centerline{\includegraphics[width=0.875\columnwidth]{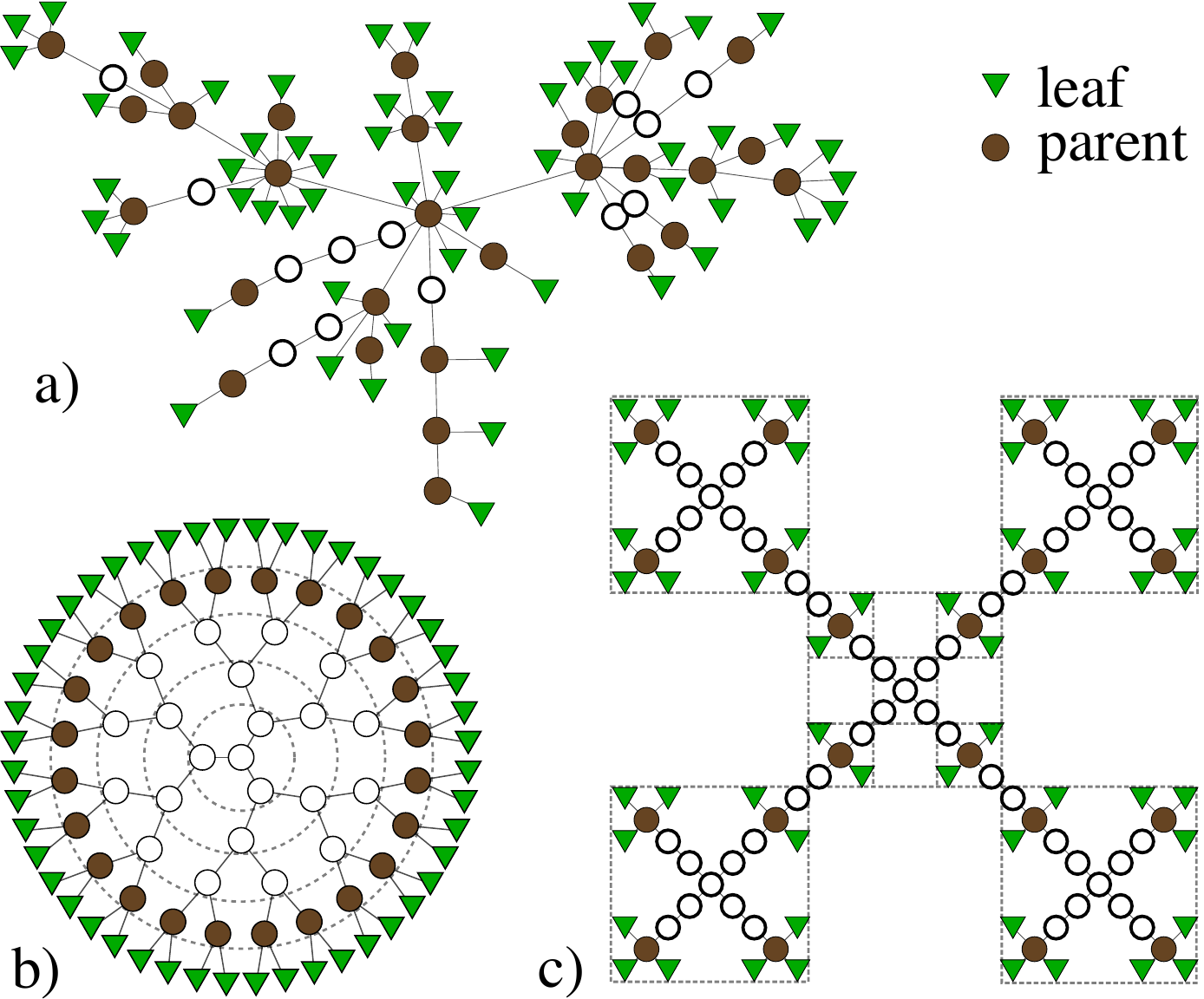}}
	\caption{(Color online) Examples of three tree-like networks: (a) a scale-free tree (SFT), where the functionalities are drawn from the probability distribution $ P(f_{j}\!=\!x)\!\propto\! x^{-s} $ (here $s=2.5$), (b) a Dendrimer (D) of generation $5$ with functionality $f=3$, and (c) a Viscek fractal (VF) of generation $3$ with functionality $f=4$. Three different types of nodes are marked: leave nodes (green triangles), parents (brown circles) of leave nodes, and all other (open circles).
}
\label{fig:graph_examples}
\end{figure}

%\section
{\sl Tree-like Networks} --
Figure \ref{fig:graph_examples} shows examples of three tree-like networks: (a) a scale-free tree (SFT), where the functionalities are drawn from the probability distribution $ P(f_{j}\!=\!x)\!\propto\! x^{-s} $ (here $s=2.5$), (b) a Dendrimer (D) of generation $5$ with functionality $f=3$, and (c) a Viscek fractal (VF) of generation $3$ with functionality $f=4$. For D, the functionality $f$ refers to those nodes having more than one bond, and for VF, it refers to the building blocks of the fractal in each iterative step, see the squares as guide to the eye in Fig.~\ref{fig:graph_examples}(c).  We do not distinguish between the respective functionalities, but we will always note by superscripts D and VF to which structure we refer to. D and VF allow for a direct computation of $\chi$ and $\underline\chi$ based on Eq.~(\ref{eq:chi_spec_dens_ineq}), since their spectra are exactly known \cite{cai1997rouse,*gotlib2002theory,jayanthi1994dynamics,Blumen2004}.  For SFT one has to resort to either numeric calculations of the spectra or to an analytic estimation of the spectral density of that eigenvalue with the highest degeneracy by counting those nodes (the \emph{leaves}, green triangles in Fig.~\ref{fig:graph_examples}) of the network which only have a single bond. In doing so, we will also need to estimate the number of nodes (the \emph{parents}, brown circles in Fig.~\ref{fig:graph_examples}) to which the leaves are connected, see below.

As is easily shown, if a parent node $j$ has two different leaves $l$ and $k$, then the superposition state $(\ket{l}-\ket{k})/\sqrt{2} $ is a normalized eigenstate of the network. The total amount of these superposition states - all belonging to the \emph{same} eigenvalue $E_*=H(f_j=1)=1$, regardless their parents because we have assumed equal couplings in $\bm H$ - can be directly calculated. 
We note, that there could be also other eigenstates, not only involving leaves and parents, leading to the same eigenvalue \cite{Blumen2004,furstenberg2013dynamics}. However, since the number of these latter states is small we concentrate on the former states.
Let $ N_{L,j} $ be the number of leaves of a single parent node $j$, then there are exactly ${N_{L,j}}-1 $ independent eigenstates of such superposition type. Clearly, the more leaves a parent node has, the larger will be the number of independent eigenstates for a given eigenvalue, i.e., the larger will be the corresponding spectral density
$\varrho(E_*) \geq (N_{\text{L}}-N_{\text{P}})/{N}$,
where $ N_{L} $ is the number of \emph{all} leaves and $ N_{\text{P}} $ is the number of \emph{all} parent nodes, which in turn can be related to $ N_{L} $: Assuming that a parent node $j$ has functionality $f_j$ and that this node is also connected to $\delta_j+1$ nodes which are not leaves, then the number of leaves of this node is $(f_j-\delta_j-1)$. Let $I_{\text{P}}$ be the index set of all parent nodes, then ${N_{\text{L}}}=\sum_{j\in I_{\text{P}}}(f_{j}-\delta_{j}-1)$. With $ N_{\text{P}} = \sum_{j\in I_{\text{P}}} 1 >1$, we obtain
\begin{equation}
\label{eq:body_surf_size}
{N_{\text{P}}}=\frac{{N_{\text{L}}}}{\avg{f-\delta}{N_\text{P}}-1}%\geq\frac{{N_{\text{L}}}}{\avg{f}{N_\text{P}}-1}\;,
\end{equation}
where $ \avg{f-\delta}{N_\text{P}} \equiv \sum_{j\in I_{\text{P}}}(f_{j}-\delta_{j})/N_{\text{P}}$ denote the average over the set of all parents. 
Given the tree-like topology, we can estimate $N_{\text L}$ based on $N$ and the average functionality of those nodes which are not leaves, $\avg{f}{N\backslash N_{\text{L}}}$: For trees there are $N-1$ bonds connecting the $N$ nodes. In the two extreme cases of chains ($\avg{f}{N\backslash N_{\text{L}}}=2$) and stars ($\avg{f}{N\backslash N_{\text{L}}}=N-1$), one has $2$ and $N-1$ leaves, respectively. For other trees the number of leaves 
\begin{equation}
	\label{eq:surf_size_2}
	{N_{\text{L}}}=N-\frac{N-2}{\avg{f}{N\backslash N_{\text{L}}}-1}\;,
\end{equation}
which lies in the interval $[2,N-1]$.
We are now in the position of expressing $\varrho(E_*)$ in terms of $N$ and the averaged functionalities  $ \avg{f}{N_{P}} $ and $ \avg{f}{N\backslash N_{\text{L}}} $.
Inserting into the RHS of Eq.~(\ref{eq:chi_spec_dens_ineq}), one obtains as a lower bound
the final result up to order $1/N$:
\begin{eqnarray}
	\label{eq:chi_res}
	\underline{\chi} &\geq& \Bigg(1-\frac{1}{\avg{f}{N\backslash N_{\text{L}}}-1}\Bigg)^2 \Bigg(1-\frac{1}{\avg{f-\delta}{N_\text{P}}-1}\Bigg)^2 \nonumber \\
&&+\frac{1}{N} 
\Bigg[ 1 - 
\Bigg(\frac{\avg{f}{N\backslash N_{\text{L}}}-2}{\avg{f}{N\backslash N_{\text{L}}}-1} \Bigg)
\Bigg(\frac{\avg{f-\delta}{N_\text{P}}-2}{\avg{f-\delta}{N_\text{P}}-1} \Bigg) \nonumber \\
&&+ 4 \Bigg(\frac{\avg{f}{N\backslash N_{\text{L}}}-2}{(\avg{f}{N\backslash N_{\text{L}}}-1)^2}\Bigg) \Bigg(\frac{\avg{f-\delta}{N_\text{P}}-2}{\avg{f-\delta}{N_\text{P}}-1}\Bigg)^2\Bigg] \;.
\end{eqnarray}
In the limit when $\avg{\delta}{N_{P}}/\avg{f}{N_{P}}\ll1$, i.e., when a parent node is only rarely coupled to more than one other parent node, one has $\avg{f-\delta}{N_{P}} = \avg{f}{N_{P}}$, where $\avg{f}{N_{P}}$ can often be written as a function of $\avg{f}{N\backslash N_{\text{L}}}$. 
If the functionality of non-leave nodes does not systematically depend on the position in the network, one has $\avg{f}{N_{P}} = \avg{f}{N\backslash N_{\text{L}}} \equiv \avg{f}{}$. 
We note, that there are exceptions, see also the VF below. Equation~(\ref{eq:chi_res}) defines only a lower bound to $\underline\chi$ because we have neglected those eigenstates which are not simple superpositions of two states localized at leaves belonging to the same parent. In the infinite system size limit, these states are negligible close to the transition point, such that the equality holds for $\underline\chi_\infty$.

Considering the inverse of the average functionality as the network's adjustable parameter, i.e., $\sigma=1/\avg{f}{}$, we can deduce a universal behavior at the breakdown of quantum transport. 
In the limit $N\to\infty$ we obtain 
\begin{equation}
\underline\chi_\infty = \left(1- \frac{1}{\avg{f}{} - 1}\right)^4, 
\end{equation}
which results to first order in $1/\avg{f}{}$ in $1-\underline\chi_\infty = 4/\avg{f}{}$ and we get
\begin{equation}
	\label{eq:univ_crit_exp}
	\underline{\kappa}=\lim\limits_{1/\avg{f}{}\to 0} \frac{\log [1-\underline\chi_\infty]}{\log [1/\avg{f}{}]}
=1\;,
\end{equation}
regardless of the original underlying network, be it deterministic or random. Therefore, all tree-like networks will yield the same \emph{universal} exponent $\underline\kappa$. We note, that the presence of loops could eventually lead to different exponents.

\begin{figure}[H]
\centerline{\includegraphics[width=0.9\columnwidth]{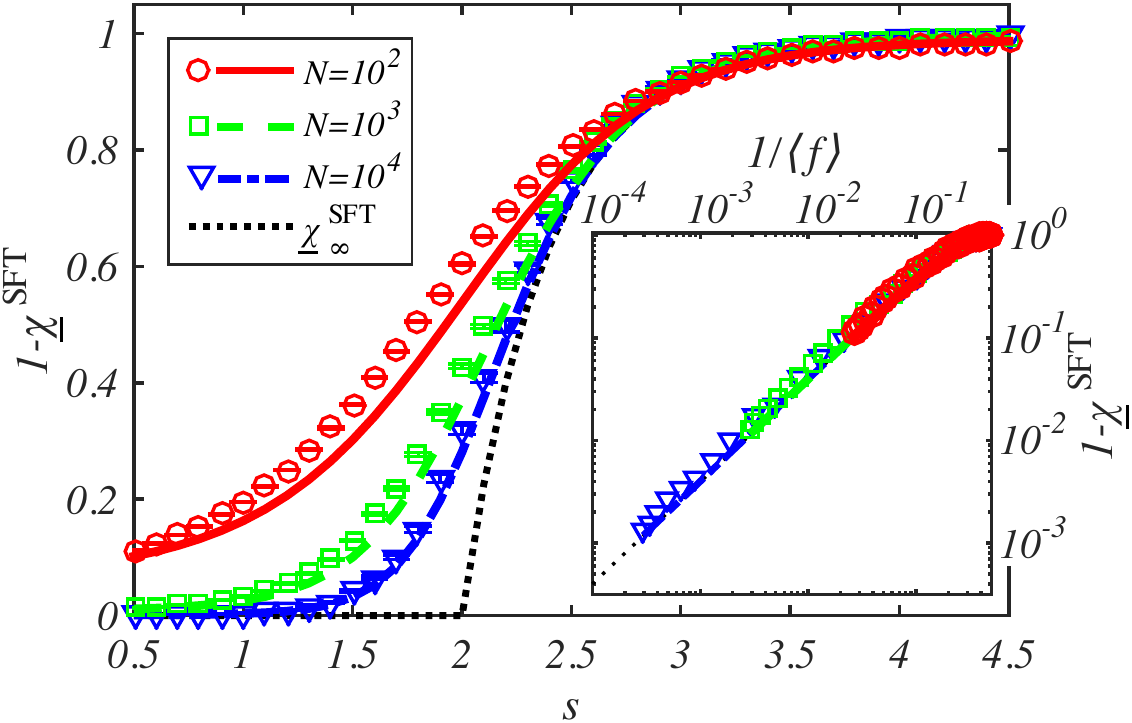}}
	\caption{(Color online) Numerical (symbols) and analytical (lines) results for $1-\underline{\chi}^{\text{SFT}}$: For finite $N=10^2, 10^3$, and $10^4$, the ensemble averages for $1-\underline{\chi}^{\text{SFT}}$ with the connectivity matrix defining $\bm H$ and with $R\!=\!10^6/N $ realizations are shown (in linear scale) as functions of the scale free parameter $s$. The analytical estimates are obtained from Eq.~(\ref{eq:chi_res}) with $\delta=0$ for finite SFT and, in the limit $N\to\infty$, from Eq.~(\ref{eq:chi_limit}). The inset shows the same curves (in log-log scale) as functions of $1/\avg{f}{}$. In both plots, one notices the clear sign of the breakdown of the quantum transport, with the expected linear behavior close to the transition at $s\searrow2$ or, equivalently, $1/\avg{f}{}\to0$.}
	\label{fig:sft}
\end{figure}

%\section
{\sl Examples} -- In order to corroborate our general findings, we consider the three examples of tree-like networks depicted in Fig.~\ref{fig:graph_examples}. All these networks allow for a parametrized transition from chain-like to star-like topologies, depending on the (average) functionality.
While D and VF allow for a direct computation of $\chi$ and $\underline\chi$ based on Eq.~(\ref{eq:chi_spec_dens_ineq}), we will employ Eq.~(\ref{eq:chi_res}) for SFT, which, depending on the average functionality, can have many leaves.
For normalisation purpose it is inevitable  for finite systems to impose a maximal functionality $f_{\max}\leq N-1$ such that the average becomes $ \avg{f}{}=\Sigma_{f=2}^{f_{\max}}f^{-s+1}/\Sigma_{f=2}^{f_{\max}}f^{-s} $.
Inserting this average in Eq.~(\ref{eq:chi_res}) yields $\underline\chi^{\text{SFT}}$. We note that $\avg{f}{}$ depends on the scaling parameter $s$ and that $1/\avg{f}{}\to0$ when $s\searrow2$. In limit $N\to\infty$ the averages are related to the Riemann zeta function $ \zeta(s)\!=\!\Sigma_{f=1}^{\infty}f^{-s} $ allowing to write the leading terms for values of $s\gtrsim2$ as
\begin{equation}
	\label{eq:chi_limit}
	\underline{\chi}_{\infty}^{\text{SFT}}
	=1-4\dfrac{\zeta(s)-1}{\zeta(s-1)-\zeta(s)}\;.
\end{equation}
Therefore, the critical exponent follows as
\begin{equation}
	\label{eq.kappa_sft}
	\underline\kappa = \lim_{s\searrow2} \frac{\log(1-\underline\chi^{\text{SFT}})}{\log(s-2)} = 1 \;,
\end{equation}
which again confirms our general statement about the universal behavior of tree-like networks. Figure~\ref{fig:sft} shows the dependence of $1-\underline\chi^{\text{SFT}}$ on the scale-free parameter $s$ and, as inset, on $\avg{f}{}$ for different sizes $N$ and also for $N\to\infty$. For finite SFT, we have compared our analytic estimation given by Eq.~(\ref{eq:chi_res}) (lines) with Monte-Carlo simulations (symbols) for SFT grown iteratively by the algorithm given in \cite{galiceanu2012relaxation} with the connectivity matrix defining $\bm H$. In the numerical computations we have considered ensemble averages of $\underline\chi^{\text{SFT}}$ with ensembles sizes of  $ R\!=\!10^6 /N $. All curves show the expected scaling close to the transition point where $1/\avg{f}{}\to0$ and $s\searrow2$, respectively. One notes from the inset of Fig.~\ref{fig:sft} that with increasing $N$ the finite-size effects become less pronounced, leading eventually to a sharp transition for $s\searrow2$. 

For D and VF, the
exact knowledge of their spectral densities \cite{cai1997rouse,*gotlib2002theory,jayanthi1994dynamics,Blumen2004} allows us to calculate $\chi$ as well as its lower bound $\underline\chi$ as functions of the respective functionalities.
According to Eq.~(\ref{eq:chi_spec_dens_ineq}), we get in the infinite size limit 
\begin{equation}
	\label{eq:dendrimer_vicsek}
	\chi^{\text{D}}_{\infty}=\left(1-\frac{2}{f}\right)^2\;\text{ and }\;
	\chi^{\text{VF}}_\infty = 1 - 6\dfrac{f-1}{f(f+2)-2}
\end{equation}
as well as the corresponding lower bounds
\begin{equation}
	\label{eq:dendrimer_vicsek_lower-bound}
	\underline\chi^{\text{D}}_{\infty}=\Bigg(1-\frac{1}{f-1}\Bigg)^4\;\text{ and }\;
	\underline\chi^{\text{VF}}_\infty = \left( 1 - \dfrac{4f-5}{f^2-1} \right)^2
\end{equation}

In both cases we find the breakdown of transport in the limit $1/f\to0$. In order to be comparable to the SFT, we express both, $\chi^{\text{D}}_{\infty}$ and $\chi^{\text{VF}}_{\infty}$, as functions of $\avg{f}{}$. For D, $\avg{f}{N\backslash N_{\text{L}}}=f$, which is not the case for VF, where $\avg{f}{N\backslash N_{\text{L}}}=(f+4)/3$ while $\avg{f}{N_\text{P}} = f$. Inserting into Eq.~(\ref{eq:dendrimer_vicsek}), we obtain for $1/\avg{f}{} \to 0$ that $1-\underline\chi^{\text{D}}_{\infty} \sim 1/\avg{f}{}$ and that $1-\underline\chi^{\text{VF}}_{\infty} \sim 1/\avg{f}{}$. Thus, also here we find the breakdown leading to the (exact) exponent $\kappa = \underline\kappa = 1$.

We finally want to stress the differences between D and VF on the one hand and SFT on the other hand: For all structures, the transition happens when $1/f\to0$ or $1/\avg{f}{} \to 0$. However, for D and VF, having \emph{fixed} deterministic functionalities $f$, there is no parameter allowing to study the behavior of $\chi_\infty$ beyond the critical point. Moreover, the limit $f\to\infty$ seems rather artificial in the $N\to\infty$ limit since no real system can ever reach this. The situation is different for SFT: Even though $\avg{f}{}$ diverges for scaling parameters $s\leq2$, it is possible to study the behavior of $\underline\chi_\infty^{\text{SFT}}$ in this parameter region. We further note that for finite SFT, one observes the maximal values of $\underline\chi^{\text{SFT}} < 1$ only in the limit when $s\to0$.

%\section
{\sl Conclusions} --
We have shown, that the breakdown of quantum transport on complex tree-like networks shows a universal behavior characterized by a global transport efficiency measure based on the time-averaged return probability. Parametrizing the corresponding Hamiltonian by the (average) functionality of the nodes of the network allows to derive bounds for this measure, leading in the infinite system size limit to a characteristic universal exponent for all tree-like networks. 
We anticipate that a similar treetment might also be feasible for other networks, also including loops. 

{\sl Acknowledgements} -- 
We thank Alex Blumen and Anastasiia Anishchenko for fruitful discussion and valuable comments. Support from the Deutsche For\-schungs\-ge\-mein\-schaft is acknowledged by OM (DFG Grant No. MU2925/1-1) and by MD (DFG Grant No. BL 142/11-1 and IRTG “Soft Matter Science” Grant No. GRK 1642/1).

%\bibliography{bib.bib}

\begin{thebibliography}{15}%
\makeatletter
\providecommand \@ifxundefined [1]{%
 \@ifx{#1\undefined}
}%
\providecommand \@ifnum [1]{%
 \ifnum #1\expandafter \@firstoftwo
 \else \expandafter \@secondoftwo
 \fi
}%
\providecommand \@ifx [1]{%
 \ifx #1\expandafter \@firstoftwo
 \else \expandafter \@secondoftwo
 \fi
}%
\providecommand \natexlab [1]{#1}%
\providecommand \enquote  [1]{``#1''}%
\providecommand \bibnamefont  [1]{#1}%
\providecommand \bibfnamefont [1]{#1}%
\providecommand \citenamefont [1]{#1}%
\providecommand \href@noop [0]{\@secondoftwo}%
\providecommand \href [0]{\begingroup \@sanitize@url \@href}%
\providecommand \@href[1]{\@@startlink{#1}\@@href}%
\providecommand \@@href[1]{\endgroup#1\@@endlink}%
\providecommand \@sanitize@url [0]{\catcode `\\12\catcode `\$12\catcode
  `\&12\catcode `\#12\catcode `\^12\catcode `\_12\catcode `\%12\relax}%
\providecommand \@@startlink[1]{}%
\providecommand \@@endlink[0]{}%
\providecommand \url  [0]{\begingroup\@sanitize@url \@url }%
\providecommand \@url [1]{\endgroup\@href {#1}{\urlprefix }}%
\providecommand \urlprefix  [0]{URL }%
\providecommand \Eprint [0]{\href }%
\providecommand \doibase [0]{http://dx.doi.org/}%
\providecommand \selectlanguage [0]{\@gobble}%
\providecommand \bibinfo  [0]{\@secondoftwo}%
\providecommand \bibfield  [0]{\@secondoftwo}%
\providecommand \translation [1]{[#1]}%
\providecommand \BibitemOpen [0]{}%
\providecommand \bibitemStop [0]{}%
\providecommand \bibitemNoStop [0]{.\EOS\space}%
\providecommand \EOS [0]{\spacefactor3000\relax}%
\providecommand \BibitemShut  [1]{\csname bibitem#1\endcsname}%
\let\auto@bib@innerbib\@empty
%</preamble>
\bibitem [{\citenamefont {Barab{\'a}si}\ \emph {et~al.}(2009)\citenamefont
  {Barab{\'a}si} \emph {et~al.}}]{barabasi2009scale}%
  \BibitemOpen
  \bibfield  {author} {\bibinfo {author} {\bibfnamefont {A.-L.}\ \bibnamefont
  {Barab{\'a}si}},\ }\href@noop {} {\bibfield  {journal}
  {\bibinfo  {journal} {Science}\ }\textbf {\bibinfo {volume} {325}},\ \bibinfo
  {pages} {412} (\bibinfo {year} {2009})}\BibitemShut {NoStop}%
\bibitem [{\citenamefont {Newman}(2010)}]{newman2010networks}%
  \BibitemOpen
  \bibfield  {author} {\bibinfo {author} {\bibfnamefont {M.}~\bibnamefont
  {Newman}},\ }\href@noop {} {\emph {\bibinfo {title} {Networks: an
  introduction}}}\ (\bibinfo  {publisher} {Oxford University Press},\ \bibinfo
  {year} {2010})\BibitemShut {NoStop}%
\bibitem [{\citenamefont {Barth{\'e}lemy}(2011)}]{barthelemy2011spatial}%
  \BibitemOpen
  \bibfield  {author} {\bibinfo {author} {\bibfnamefont {M.}~\bibnamefont
  {Barth{\'e}lemy}},\ }\href@noop {} {\bibfield  {journal} {\bibinfo  {journal}
  {Phys.\ Rep.}\ }\textbf {\bibinfo {volume} {499}},\ \bibinfo {pages} {1}
  (\bibinfo {year} {2011})}\BibitemShut {NoStop}%
\bibitem [{\citenamefont {Albert}\ \emph {et~al.}(1999)\citenamefont {Albert},
  \citenamefont {Jeong},\ and\ \citenamefont {L.}}]{albert1999internet}%
  \BibitemOpen
  \bibfield  {author} {\bibinfo {author} {\bibfnamefont {R.}~\bibnamefont
  {Albert}}, \bibinfo {author} {\bibfnamefont {H.}~\bibnamefont {Jeong}}, \
  and\ \bibinfo {author} {\bibfnamefont {A.-L.}\ \bibnamefont {Barab{\'a}si}},\
  }\href@noop {} {\bibfield  {journal} {\bibinfo  {journal} {Nature}\ }\textbf
  {\bibinfo {volume} {401}},\ \bibinfo {pages} {130} (\bibinfo {year}
  {1999})}\BibitemShut {NoStop}%
\bibitem [{\citenamefont {Gurtovenko}\ and\ \citenamefont
  {Blumen}(2005)}]{Gurtovenko2005}%
  \BibitemOpen
  \bibfield  {author} {\bibinfo {author} {\bibfnamefont {A.~A.}\ \bibnamefont
  {Gurtovenko}}\ and\ \bibinfo {author} {\bibfnamefont {A.}~\bibnamefont
  {Blumen}},\ }\href@noop {} {\bibfield  {journal} {\bibinfo  {journal} {Adv.
  Polym. Sci.}\ }\textbf {\bibinfo {volume} {182}},\ \bibinfo {pages}
  {171–282} (\bibinfo {year} {2005})}\BibitemShut {NoStop}%
\bibitem [{\citenamefont {Reuveni}\ \emph {et~al.}(2010)\citenamefont
  {Reuveni}, \citenamefont {Granek},\ and\ \citenamefont
  {Klafter}}]{Reuveni2010}%
  \BibitemOpen
  \bibfield  {author} {\bibinfo {author} {\bibfnamefont {S.}~\bibnamefont
  {Reuveni}}, \bibinfo {author} {\bibfnamefont {R.}~\bibnamefont {Granek}}, \
  and\ \bibinfo {author} {\bibfnamefont {J.}~\bibnamefont {Klafter}},\
  }\href@noop {} {\bibfield  {journal} {\bibinfo  {journal} {Proc. Natl. Acad.
  Sci. USA}\ }\textbf {\bibinfo {volume} {107}},\ \bibinfo {pages} {13696}
  (\bibinfo {year} {2010})}\BibitemShut {NoStop}%
\bibitem [{\citenamefont {Barzel}\ and\ \citenamefont
  {Barab{\'a}si}(2013)}]{barzel2013universality}%
  \BibitemOpen
  \bibfield  {author} {\bibinfo {author} {\bibfnamefont {B.}~\bibnamefont
  {Barzel}}\ and\ \bibinfo {author} {\bibfnamefont {A.-L.}\ \bibnamefont
  {Barab{\'a}si}},\ }\href@noop {} {\bibfield  {journal} {\bibinfo  {journal}
  {Nat.\ Phys.}\ }\textbf {\bibinfo {volume} {9}},\ \bibinfo {pages} {673}
  (\bibinfo {year} {2013})}\BibitemShut {NoStop}%
\bibitem [{\citenamefont {M{\"u}lken}\ and\ \citenamefont
  {Blumen}(2011)}]{mulken2011continuous}%
  \BibitemOpen
  \bibfield  {author} {\bibinfo {author} {\bibfnamefont {O.}~\bibnamefont
  {M{\"u}lken}}\ and\ \bibinfo {author} {\bibfnamefont {A.}~\bibnamefont
  {Blumen}},\ }\href@noop {} {\bibfield  {journal} {\bibinfo  {journal}
  {Phys.\ Rep.}\ }\textbf {\bibinfo {volume} {502}},\ \bibinfo {pages} {37}
  (\bibinfo {year} {2011})}\BibitemShut {NoStop}%
\bibitem [{\citenamefont {Farhi}\ and\ \citenamefont
  {Gutmann}(1998)}]{farhi1998quantum}%
  \BibitemOpen
  \bibfield  {author} {\bibinfo {author} {\bibfnamefont {E.}~\bibnamefont
  {Farhi}}\ and\ \bibinfo {author} {\bibfnamefont {S.}~\bibnamefont
  {Gutmann}},\ }\href@noop {} {\bibfield  {journal} {\bibinfo  {journal}
  {Phys.\ Rev.\ A}\ }\textbf {\bibinfo {volume} {58}},\ \bibinfo {pages}
  {915} (\bibinfo {year} {1998})}\BibitemShut {NoStop}%
\bibitem [{\citenamefont {Cai}\ and\ \citenamefont
  {Chen}(1997)}]{cai1997rouse}%
  \BibitemOpen
  \bibfield  {author} {\bibinfo {author} {\bibfnamefont {C.}~\bibnamefont
  {Cai}}\ and\ \bibinfo {author} {\bibfnamefont {Z.~Y.}\ \bibnamefont {Chen}},\
  }\href@noop {} {\bibfield  {journal} {\bibinfo  {journal} {Macromolecules}\
  }\textbf {\bibinfo {volume} {30}},\ \bibinfo {pages} {5104} (\bibinfo {year}
  {1997})}\BibitemShut {NoStop}%
\bibitem [{\citenamefont {Gotlib}\ and\ \citenamefont
  {Markelov}(2002)}]{gotlib2002theory}%
  \BibitemOpen
  \bibfield  {author} {\bibinfo {author} {\bibfnamefont {Y.~Y.}\ \bibnamefont
  {Gotlib}}\ and\ \bibinfo {author} {\bibfnamefont {D.~A.}\ \bibnamefont
  {Markelov}},\ }\href@noop {} {\bibfield  {journal} {\bibinfo  {journal}
  {Polym. Sci. Ser. A}\ }\textbf {\bibinfo {volume} {44}},\ \bibinfo {pages}
  {1341} (\bibinfo {year} {2002})}\BibitemShut {NoStop}%
\bibitem [{\citenamefont {Jayanthi}\ and\ \citenamefont
  {Wu}(1994)}]{jayanthi1994dynamics}%
  \BibitemOpen
  \bibfield  {author} {\bibinfo {author} {\bibfnamefont {C.~S.}\ \bibnamefont
  {Jayanthi}}\ and\ \bibinfo {author} {\bibfnamefont {S.~Y.}\ \bibnamefont
  {Wu}},\ }\href@noop {} {\bibfield  {journal} {\bibinfo  {journal} {Phys.\
  Rev.\ B}\ }\textbf {\bibinfo {volume} {50}},\ \bibinfo {pages} {897}
  (\bibinfo {year} {1994})}\BibitemShut {NoStop}%
\bibitem [{\citenamefont {Blumen}\ \emph {et~al.}(2004)\citenamefont {Blumen},
  \citenamefont {von Ferber}, \citenamefont {Jurjiu},\ and\ \citenamefont
  {Koslowski}}]{Blumen2004}%
  \BibitemOpen
  \bibfield  {author} {\bibinfo {author} {\bibfnamefont {A.}~\bibnamefont
  {Blumen}}, \bibinfo {author} {\bibfnamefont {C.}~\bibnamefont {von Ferber}},
  \bibinfo {author} {\bibfnamefont {A.}~\bibnamefont {Jurjiu}}, \ and\ \bibinfo
  {author} {\bibfnamefont {T.}~\bibnamefont {Koslowski}},\ }\href@noop {}
  {\bibfield  {journal} {\bibinfo  {journal} {Macromolecules}\ }\textbf
  {\bibinfo {volume} {37}},\ \bibinfo {pages} {638} (\bibinfo {year}
  {2004})}\BibitemShut {NoStop}%
\bibitem [{\citenamefont {F{\"u}rstenberg}\ \emph {et~al.}(2013)\citenamefont
  {F{\"u}rstenberg}, \citenamefont {Dolgushev},\ and\ \citenamefont
  {Blumen}}]{furstenberg2013dynamics}%
  \BibitemOpen
  \bibfield  {author} {\bibinfo {author} {\bibfnamefont {F.}~\bibnamefont
  {F{\"u}rstenberg}}, \bibinfo {author} {\bibfnamefont {M.}~\bibnamefont
  {Dolgushev}}, \ and\ \bibinfo {author} {\bibfnamefont {A.}~\bibnamefont
  {Blumen}},\ }\href@noop {} {\bibfield  {journal} {\bibinfo  {journal} {J.\ 
  Chem.\ Phys.}\ }\textbf {\bibinfo {volume} {138}},\ \bibinfo
  {pages} {034904} (\bibinfo {year} {2013})}\BibitemShut {NoStop}%
\bibitem [{\citenamefont {Galiceanu}(2012)}]{galiceanu2012relaxation}%
  \BibitemOpen
  \bibfield  {author} {\bibinfo {author} {\bibfnamefont {M.}~\bibnamefont
  {Galiceanu}},\ }\href@noop {} {\bibfield  {journal} {\bibinfo  {journal}
  {Phys.\ Rev.\ E}\ }\textbf {\bibinfo {volume} {86}},\ \bibinfo {pages}
  {041803} (\bibinfo {year} {2012})}\BibitemShut {NoStop}%
\end{thebibliography}
%\end{document}

\end{document}